# Self-supervised Model Based on Masked Autoencoders Advance CT Scans Classification

**Jiashu Xu**
National Technical University of Ukraine "Igor Sikorsky Kyiv Polytechnic Institute", Kyiv, 03056, Ukraine
E-mail: xu.jiashu@lll.kpi.ua

**Sergii Stirenko**
National Technical University of Ukraine "Igor Sikorsky Kyiv Polytechnic Institute", Kyiv, 03056, Ukraine
E-mail: stirenko@comsys.kpi.ua

**Abstract:** The coronavirus pandemic has been going on since the year 2019, and the trend is still not abating. Therefore, it is particularly important to classify medical CT scans to assist in medical diagnosis. At present, Supervised Deep Learning algorithms have made a great success in the classification task of medical CT scans, but medical image datasets often require professional image annotation, and many research datasets are not publicly available. To solve this problem, this paper is inspired by the self-supervised learning algorithm MAE and uses the MAE model pre-trained on ImageNet to perform transfer learning on CT Scans dataset. This method improves the generalization performance of the model and avoids the risk of overfitting on small datasets. Through extensive experiments on the COVID-CT dataset and the SARS-CoV-2 dataset, we compare the SSL-based method in this paper with other state-of-the-art supervised learning-based pretraining methods. Experimental results show that our method improves the generalization performance of the model more effectively and avoids the risk of overfitting on small datasets. The model achieved almost the same accuracy as supervised learning on both test datasets. Finally, ablation experiments aim to fully demonstrate the effectiveness of our method and how it works.

**Index Terms:** Self-supervised learning, CT scans, Transfer Learning, Classification.

## 1. Introduction

Data sets in the field of the medical image generally face the following challenges: 1. Privacy issues. To protect the privacy of patients, most medical imaging data are not public. 2. There is a lack of open source annotated medical image datasets because the annotation of medical images often requires professionals to complete. Because of these problems, the development of deep learning in the medical field has been hindered to a certain extent. This paper aims to solve these problems and reduce the dependence on large sets of annotated data. Current methods to address these problems include data augmentation and transfer learning. Data augmentation improves model generalization performance by augmenting data, such as using GAN to obtain more samples based on generative models. Transfer learning improves generalization performance through models with generic feature extraction obtained in exotic domains. However, these two methods have their limitations and cannot completely solve the above problems. We are inspired by the self-supervised learning algorithm Masked Autoencoder (MAE) [1], our paper uses the strategy of masked autoencoder to pre-training the encoder and decoder [2], and then fine-tune the pre-trained encoder on CT scans dataset for transfer learning [3]. MAE is a strategy for generating (predicting) pretraining types. A well-known example of such self-supervised learning is BERT [4]. The BERT model in the natural language processing task covers some tokens in the middle of the sentence, allowing the model to predict and minimize the loss between the obtained prediction result and the real token. Through the process of reconstructing the sentence, the model learns latent representations of the data. Similarly, for the MAE model, some patches are randomly covered in an image, allowing the model to predict these patches, finally, minimize the distance between the predicted results and the real image patches, this process achieves self-supervised feature learning.

*Contributions*

1. We propose a self-supervised learning system to improve the classification task of CT scans (COVID-19). Our code will be open source.

2. We conduct extensive experiments on transfer learning and integrative studies across different models to compare the results of transfer learning with supervised and self-supervised learning for COVID-19 diagnosis, ultimately providing insightful findings.

3. Comparing our experimental results with previous work, our method achieves some progress, and our results are as follows: The experimental results show that the MAE-based method achieves almost the same performance as supervised learning, and the ablation experiments deeply analyze the efficiency of self-supervised MAE.

Related work will be reviewed in Section 2. The method of this paper is presented in Section 3, the experimental design and results are introduced in Section 4, and the paper is concluded in Section 5.

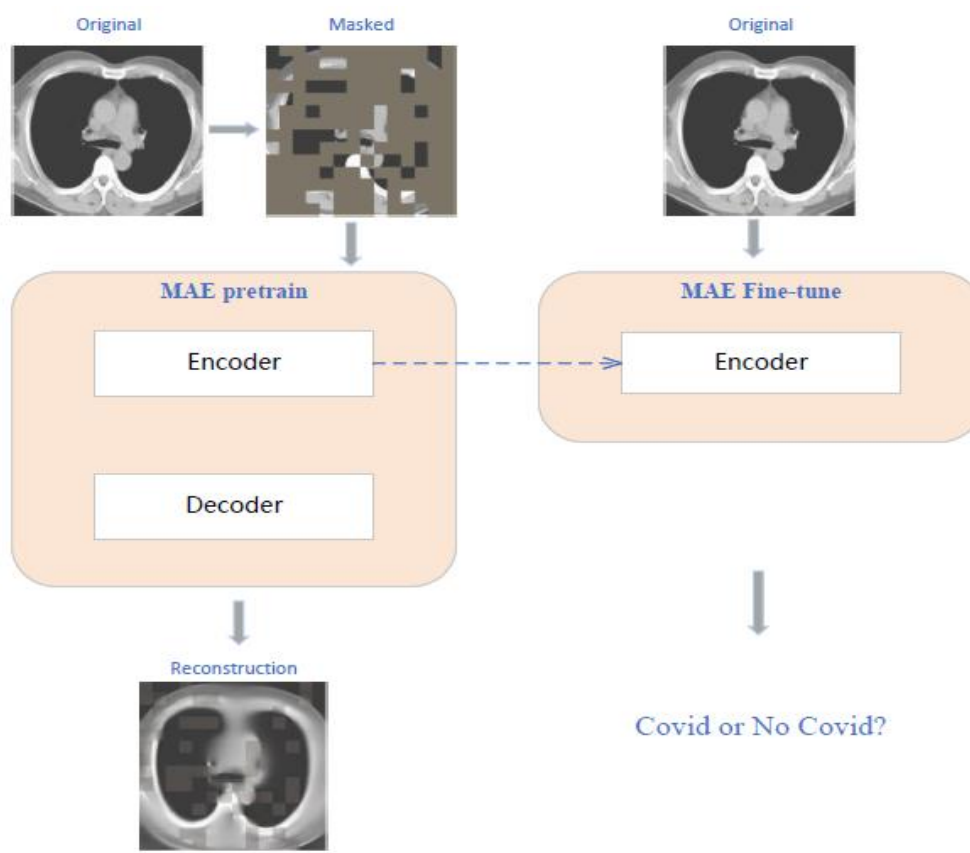

Fig. 1. MAE for medical image

## 2. Related Work

In the present two years since the COVID-19 outbreak, researchers have developed numerous deep learning-based frameworks for COVID-19 diagnosis [5,6], Vinayakumar et al. [7] proposed a deep learning-based approach for COVID-19 classification meta-classifier that uses EfficientNet [16] as a pre-trained model for feature extraction. Sun et al. [8] proposed an AFS deep forest for COVID-19 CT image classification, specifically, using a deep forest to learn high-level representations based on location-specific features. These methods are based on supervised learning and thus mainly contribute to dataset collection but have limitations in the algorithms. Anuroop et al. [9] pre-trained the self-supervised model MOCO [10] on two large public chest X-ray datasets [11,12], and the COVID-19 classification task as a downstream task, then predict COVID-19 patient outcomes. Coincidentally, He et al. [13] also borrowed the MOCO self-supervised learning algorithm and proposed a Self-Trans Pipeline for COVID-19 image classification. Li et al. [14] propose to utilize U-net [15] to reconstruct images as a self-supervised pretext task, and then use the trained feature extraction capabilities for the downstream task COVID-19 CT image classification. These MOCO-based self-supervised methods have achieved success, but MOCO is based on the idea of comparative learning, which requires a lot of computing resources, and the model efficiency needs to be improved.

### 2.1. Self-supervised learning

Self-supervised learning is a subset of unsupervised learning. The main idea of SSL is to use pre-text tasks to learn common feature representations from large-scale unsupervised data, and then obtain a pre-trained model to transfer it to new downstream tasks. According to the strategy of the pre-text tasks, self-supervised learning can be divided into three main categories: Context-Based, Contrastive Based, Generation Based [17]. Context-based methods usually do the prediction of geometric transformations, and the typical work is [18,19]. The method of contrastive learning is the current mainstream, the most important representative is MOCO and the recent work iBOT [20] introduces Vision Transformers (ViT) [21]. Generative SSL: the goal of the method is to reconstruct the input, the input and output are as similar as possible, such as GAN [22], VAE autoencoder [23]. But these methods focus on leveraging self-supervision to learn general representations regardless of the performance of downstream tasks, our work aims to improve the performance of transfer learning through self-supervised pre-training on unlabeled data.

### 2.2. Vision Transformer

Transformer first appeared as a classic NLP model proposed by Google's team in 2017 [24]. The Transformer model uses the Self-Attention mechanism and does not use the sequential structure of RNN, so that the model can be trained in parallel and can have global information. The input of the Transformer is a sequence, then in the visual task, the Transformer converts a bunch of pictures into a sequence. The specific method is to divide the image into patches, then treat each patch as a vector, and all vectors together as a sequence. The difference between a CNN and a Vision Transformer is that traditional CNNs treat all image pixels with equal importance, regardless of what the content of that pixel is and whether it is important or not. Clearly, this is unreasonable, e.g., image classification model should prioritize foreground objects over backgrounds. In addition, the weakness of convolution is that it is difficult to connect spatially distant concepts. To face the problem, it is often necessary to increase model complexity to alleviate this problem. In Vision Transformer, instead of modeling all concepts in all images, semantic concepts are encoded in visual tokens, and then Transformer is used to model the relationship between tokens. In this paper, the self-supervised learning pre-text task is constructed inspired by the architectural properties of the Vision Transformer.

## 3. Method

Inspired by the self-supervised learning algorithm Masked Autoencoders [1], our overall framework is shown in Figure 1. The self-supervised process is shown in the left half of Figure 1, this part is an asymmetric encoder-decoder, the input to the encoder is the image that randomly masked some patches of a certain ratio. The straightforward motivation for the encoder-decoder architecture is reconstructed missing patches of the original image. Specifically, the image is divided into regular, non-overlapping patches. Then some non-repeated patches are selected according to the uniform distribution, and the rest are masked out. The mask rate should be high enough. The experiment of MAE[1] pointed out that when the mask rate is 75%, the self-supervised learning effect is the best, which greatly reduces the redundant information of the patch and increases the difficulty of reconstructing the image. On the right side of Figure 1, the trained encoder is reserved for the classification of downstream tasks.

### 3.1 Encoder

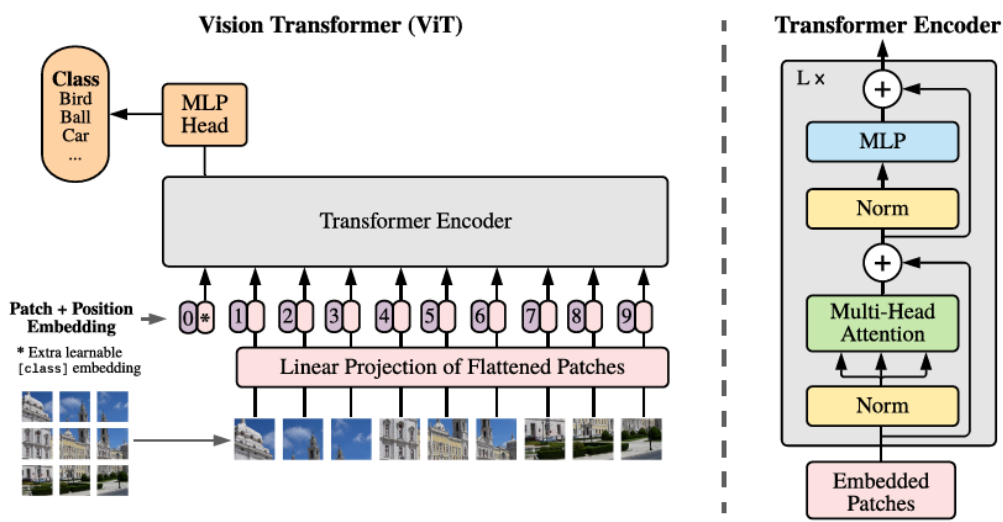

Fig. 2. ViT [21]

The encoder uses ViT architecture [25], but it only works on unmasked patches. The Encoder will first encode the unmasked patches through Linear Projection, plus position encoding, and then send it into Transformer Blocks. The ViT architecture is shown in Figure 2. The encoder here operates only on a small subset of the entire set of image patches, removing the remaining masked patches. This is different from BERT, which uses special characters to replace the masked part.

Transformer encoder part, mainly including multi-head self-attention (MSA) and Multilayer Perceptron (MLP) two parts. Mathematical modeling process of the single-layer Encoder structure of ViT:

Use X to represent the input data, and obtain three vectors of Q (query), K (key), and V (value) through linear transformation. As shown in formulas 1, 2, and 3, where PE refers to the position encode.

$$Q = W^q(X + PE) \tag{1}$$

$$K = W^k(X + PE) \tag{2}$$

$$V = W^v(X + PE) \tag{3}$$

Then, calculate the similarity between Q and K, and then multiply this similarity by V to obtain the final self-attention output, finally use a linear fully connected layer to map the intermediate feature dimension to the output feature dimension. As shown in formula 4, where LN refers to LayerNorm.

$$X = LN(linear(soft\max(\frac{QK^T}{\sqrt{d_k}})V) + X) \tag{4}$$

After completing Self-Attention, a layer of Feed Forward Network (FFN) will be added to process the X input after self-attention processing, as shown in the following formula 5.

$$Y = LN(FFN(X) + X) \tag{5}$$

### 3.2 Decoder

The Decoder performs image reconstruction tasks during self-supervised pre-training, mapping low-rank representations to reconstructions. Although the decoder is only used for pre-training, it plays a crucial role in the feature learning of the encoder. It is for this reason that the decoder structure can be flexibly designed, and tokens can be processed by a lightweight decoder. Under this asymmetric encoder-decoder design, the pre-training time is greatly shortened. The decoder also adopts the Transformer architecture, in which inputs are the set of tokens corresponding to the entire image patches, not only the unmasked tokens but also the masked ones. Each mask token is a shared learning vector indicating that there is a token to predict. Additionally, positional embeddings are added to all tokens in this complete set of image patches, where positional encodings represent information about the position of each patch in the image. The last layer of the Decoder is a Linear Projection layer, which outputs the number of channels equal to the number of pixels of the image. Therefore, the output of the Decoder will be further reshaped into the shape of the original image. Finally, the loss function directly minimizes the distance between the reconstructed image and the original image using the mean squared error (MSE).

### 3.3 Transfer learning

Transfer learning has two basic concepts: Domain and Task. It is to transfer the model parameters trained in the source domain to the target domain to help the new model train to solve the task of the target domain. A common approach is to pre-train a neural network on a large dataset in the source domain task, this process is used for feature extraction, and then fine-tune the model on the task in the target domain. In our paper, the encoder is pretrained on the ImageNet dataset [26] by a self-supervised method, and then the encoder is used for transfer learning in the CT image classification task. Using this strategy, it is necessary to consider the large domain discrepancy between the images of ImageNet and the images of CT scans. To investigate this problem, our experiments are conducted on two datasets from different domains: ImageNet and COVID-CTset [27]. The ImageNet dataset is intuitively very different from the target domain, while the COVID-CTset dataset seems to be close to the target domain. In the experiments, the effects of pre-trained models under different backbone networks on transfer learning were also investigated. The implementation and results of the experiments are described in detail in the next section.

## 4. Experiments

In this subsection, we extensively evaluate randomly initialized baselines with different backbone networks, transfer networks based on ImageNet pre-training, and the self-supervised learning method we use.

### 4.1 Datasets

**COVID-CTset** [27] is a Large COVID-19 CT Scans dataset of more than 60000 CT images. This dataset is used as pre-training for the self-supervised learning stage.

**COVID-CT-Dataset** [28] is a small data set, with 397 COVID19 negative samples and 349 COVID19 positive samples. This dataset is used in the downstream tasks of the experiment.

**SARS-CoV-2** [29] CT Scan Dataset, the data set is published on the Kaggle platform, with a total sample size of 2482. The positive and negative samples in the data set are approximately balanced.

### 4.2 Experimental environment

These models are implemented using the deep learning framework PyTorch. Use GPU-accelerated training on the Google Colab pro platform. Data augmentation is implemented in the dataset preprocessing stage to improve the generalization ability of these models.

### 4.3 Evaluate with random initialized models

To demonstrate the effectiveness of the self-supervised learning approach, we first conduct experiments on randomly initialized models, using different backbone networks. Including ResNet-101 [30], DenseNet-121[31], VGG-16 [32], EfficientNetb0 [33], EfficientNet-b1 and our proposed SSL.

Experimental implementation details: Classifiers trained from scratch, these models are trained with 1000 epochs on the COVID-CT-Dataset and 500 epochs on the SARS-CoV-2. The optimizer uses Adam with initial learning rate is 0.0001, and apply the cosine annealing learning rate varies periodically, the 1/2 of the period is 10.

Table 1 shows the training results of the initialized model on the two datasets. The experimental results show that the EfficientNet model has better results. This is mainly because EfficientNet can balance the depth, width, and resolution of the model for better performance, and uses a compound coefficient to uniformly adjust the scale of the model. Figure 3 shows the change curve of the average accuracy rate of the test dataset during the training process. Due to the small size of the test set and the random initialization of the model training, the acc curve oscillates in a small range. The purpose of setting the initialization experiment is to compare with the next experiments in 4.4 and 4.5.

Table 1. Performance comparison of random initialization of different backbone networks on two downstream task datasets

| Networks | Acc | | F1 | | AUC | |
|---|---|---|---|---|---|---|
| | COVID-CT | SARS-CoV-2 | COVID-CT | SARS-CoV-2 | COVID-CT | SARS-CoV-2 |
| ResNet-101 | 0.8136 | 0.9756 | 0.8167 | 0.9744 | 0.9152 | 0.9990 |
| DenseNet-121 | 0.8136 | 0.9725 | 0.8305 | 0.9722 | 0.9046 | 0.9966 |
| VGG-16 | 0.8051 | 0.9750 | 0.8000 | 0.9749 | 0.9026 | 0.9962 |
| EfficientNetb0 | 0.8220 | 0.9798 | 0.8293 | **0.9824** | 0.9089 | **0.9992** |
| EfficientNet-b1 | **0.8290** | **0.9825** | **0.8380** | 0.9800 | **0.9155** | 0.9977 |

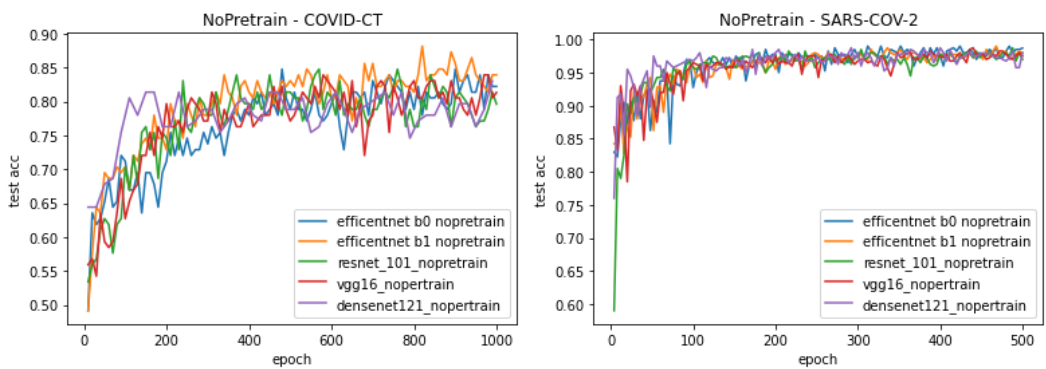

Fig. 3. The evaluation results of different models trained with random initialization, on the left are the evaluation results on dataset COVID-CT, and the right evaluation results on dataset SARS-CoV-2.

### 4.4 Transfer Learning Based on Supervised Pretraining

To evaluate the performance of different backbone networks under transfer learning, in this experimental phase, these models in Section 4.3 are pre-trained on the large-scale dataset ImageNet, then fine-tuned on two downstream tasks.

Table 2. Performance comparison between SSL-MAE model (ours.) And IMAGENET pretrained models

| Networks | Acc | | F1 | | AUC | |
|---|---|---|---|---|---|---|
| | COVID-CT | SARS-CoV-2 | COVID-CT | SARS-CoV-2 | COVID-CT | SARS-CoV-2 |
| ResNet-101 | 0.8729 | **0.9900** | 0.8739 | 0.9899 | 0.9537 | 0.9993 |
| DenseNet-121 | 0.8644 | 0.9850 | 0.8667 | 0.9848 | 0.9534 | **0.9997** |
| VGG-16 | 0.8814 | 0.9800 | 0.8852 | 0.9797 | 0.9557 | 0.9992 |
| EfficientNetb0 | 0.8814 | 0.9875 | 0.8772 | 0.9875 | **0.9681** | 0.9996 |
| EfficientNet-b1 | **0.9068** | **0.9900** | **0.9076** | **0.9900** | 0.9549 | 0.9994 |
| MAE (ours) | 0.8898 | 0.9875 | 0.8908 | 0.9875 | 0.9637 | 0.9996 |

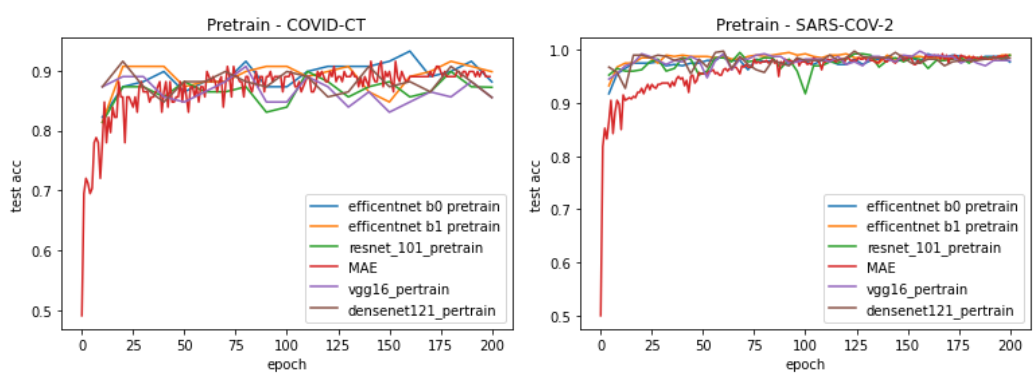

Fig. 4. Evaluation results of different models for transfer training, the left is the evaluation result of the dataset COVID-CT, and the right is the evaluation result of the dataset SARS-CoV-2.

Comparing the experimental results in Table 1 and Table 2, the results after pre-training on ImageNet are greatly improved compared to the random initialization model. For example, the EfficientNetb0 model trained with random initialization on the dataset COVID-CT obtained an average acc of 82.20%, F1 of 82.93%, and AUC of 90.89%. With transfer learning, acc is improved by 6% (absolute), F1 is improved by 5% (absolute), and AUC is improved by 6% (absolute), the training results of these models on the dataset SARS-CoV-2 also demonstrate the improvement of model performance by transfer learning.

In addition, by comparing Figure 3 and Figure 4, the oscillation of the training curve of the pre-trained model becomes smoother, the training process becomes stable, and the convergence speed in the early stage of training becomes faster. This is mainly due to transfer learning, which enables the model to learn general visual feature extraction capabilities. The essence of the self-supervised learning algorithm is also a kind of transfer learning, but the self-supervised learning algorithm is more pre-trained in the same data domain as the target domain, while the supervised transfer learning generally does not need to be pre-trained in the same data domain. The method based on self-supervised not only has the advantages of transfer learning to obtain general data extraction ability but also can ignore the label of the data.

### 4.5 Evaluate Self-Supervised Learning MAE

The details of the experiments in the self-supervised training phase are set up as follows: The decoder also uses Vision Transformer as the backbone network, with a depth of 8 and a width of 512, and the reconstructed pixels are non-normalized. The data augmentation strategy uses the usual random cropping and random resizing. The input image is reshaped to 224*224 and then divided into patches with size 16*16, the end 75% of these patches are randomly masked.

As can be seen in Figure 4, the red curve is the training process curve of MAE. This method is applied to the pre-trained MAE model on the ImageNet dataset, and then fine-tune is performed on the dataset COVID-CT and SARS-CoV-2 respectively. Experimental results show that MAE-based method and the supervised learning-based methods achieve almost the same performance in downstream tasks, even better than Vgg16, resnet101, and densenet121 models. In addition, it can be seen from the experimental results that the training curve based on MAE is smoother, which indicates that the model is more stable and has better generalization performance.

To fully find out the performance of self-supervised learning MAE, we conducted the following ablation studies:

Test 1. First, initialize the weights randomly, then pre-train MAE on the COVID-CTset without using the labels of the COVID-CTset, and finally use the pre-trained model to fine-tune on the two downstream tasks respectively.

Test 2. First, initialize the weights randomly, then pre-train MAE on the ImageNet dataset without using the labels, and finally use the pre-trained model to fine-tune on the two downstream tasks respectively.

Test 3. First, use the pre-trained model (self-supervised pre-trained on the ImageNet dataset), then perform pre-training MAE on the COVID-CTset without using the labels of the COVID-CTset, and finally fine-tune on the two downstream tasks.

Test 4. First, use the pre-trained model (self-supervised pre-trained on the ImageNet dataset), then perform pre-training MAE on the COVID-CT without using the labels of the COVID-CT, and finally fine-tune on the COVID-CT.

Test 5. First, use the pre-trained model (pre-trained on the ImageNet dataset), then perform pre-training MAE on the SARS-CoV-2 without using the labels of the SARS-CoV-2, and finally fine-tune on the SARS-CoV-2.

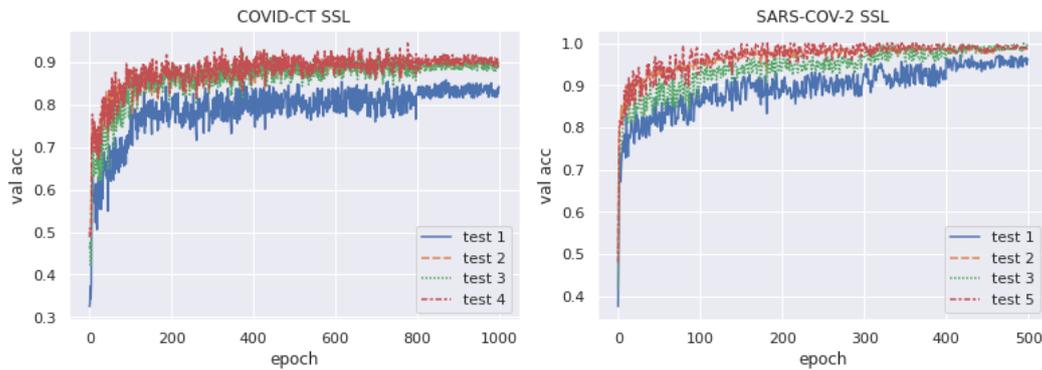

Fig. 5. Ablation study of self-supervised MAE models on these two medical image datasets

Table 3. Performance Of Mae in The Ablation Studies

| Networks | Acc | | F1 | | AUC | |
|---|---|---|---|---|---|---|
| | COVID-CT | SARS-CoV-2 | COVID-CT | SARS-CoV-2 | COVID-CT | SARS-CoV-2 |
| Test 1 | 0.8428 | 0.9633 | 0.8339 | 0.9588 | 0.9025 | 0.9996 |
| Test 2 | 0.8898 | 0.9875 | 0.8908 | 0.9823 | 0.9637 | 0.9996 |
| Test 3 | 0.8914 | 0.9870 | 0.8931 | 0.9797 | 0.9721 | 0.9998 |
| Test 4 | 0.8924 | 0.9920 | 0.8912 | 0.9875 | 0.9681 | 0.9998 |

Table 3 shows the results of the ablation studies, and Figure 5 shows the training curves of the ablation experimental models on downstream tasks. Analyzing the experimental results, we observe the following:

1. Compare test1 and test2 schemes, pre-training MAE on ImageNet outperforms pre-training on COVID-CTset. Although the image domain of COVID-CTset is closer to the target domain of downstream tasks, the experimental results are not ideal. Self-supervised learning needs to be pre-trained on large datasets to better obtain powerful feature extraction capabilities.
2. Comparing the schemes of test2 and test3, the experimental results are not much different and are relatively close. This shows that pre-training on datasets close to the target domain cannot improve the generalization performance of self-supervised models.
3. Comparing the schemes of test3, test4, and test5, it is found that pre-training in the target domain can effectively improve the performance of downstream tasks.

Transfer learning has been proven to be useful in many previous works [34,35,36,[36]. In this paper, the importance of transfer learning is again proved in experiments, and it is also shown that self-supervised learning methods can also achieve the same performance as supervised transfer learning and ignoring the data label can greatly improve the efficiency of obtaining a general model.

## 5. Conclusions

In this paper, we investigate the performance of MAE-based self-supervised learning algorithms on medical small-sample datasets. Almost all small sample datasets face the challenge of overfitting. The general traditional way is transfer learning, that is, using pre-trained models on large datasets to deal with the problem of overfitting. In this study, to compare the MAE-based self-supervised learning method and the supervised pre-training-based method, we conducted experiments on the COVID19-CT dataset and SARS-CoV-2, then found that our method is equally effective in downstream tasks, it can achieve the same effect as supervised learning.

Finally, perform ablation research, and the following conclusions were obtained: in the pre-training stage, pre-training on a data domain unrelated to the target data domain and pre-training on a data domain similar to the target data domain have the same effect, and there is not much difference. However, when the step of pre-training on the target domain is added, the performance of the model is improved. This is superior to supervised transfer learning, where self-supervised methods can improve model performance even on unlabeled target data domains. This method will help improve the efficiency of medical image diagnosis and provide some prior experience to researchers and medical image analysts.

## Acknowledgment

This research has been partially supported by China Scholarship Council (CSC), and Special thanks should go to my supervisor professor Sergii Stirenko, for his instructive advice and useful suggestions on my paper.

**Authors' Profiles**

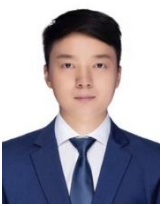

**Jiashu Xu** received a master's degree from the Department of computing engineering, National Technical University of Ukraine "Igor Sikorsky Kyiv Polytechnic Institute". Now is a Ph.D. student from the same university. His research interests include self-supervised learning, unsupervised learning, computer vision, GAN, and their applications in the medical image domain.
.

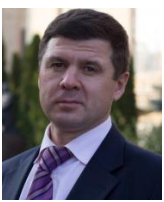

**Sergii Stirenko,** Head of Computer Engineering Department, Research Supervisor of KPI-Samsung R&D Lab, Head of NVIDIA GPU Education and NVIDIA GPU Research Center, and Professor at National Technical University of Ukraine "Kyiv Polytechnic Institute." Research is mainly focused on artificial intelligence, high-performance computing, cloud computing, distributed computing, parallel computing, eHealth, simulations, and statistical methods. And published more than 60 papers in peer-reviewed international journals.